\documentstyle[12pt]{article} 
\newcommand{\be}{\begin{equation}} 
\newcommand{\ee}{\end{equation}} 
\newcommand{\bea}{\begin{eqnarray}} 
\newcommand{\eea}{\end{eqnarray}}

\begin{document} 
 
\title{Topology of Equivalent  Unconstrained Systems in QCD} 
 
\author{Victor Pervushin\\
BLTP, Joint Institute for Nuclear Research, 141980 Dubna, Russia.} 
 
 
\maketitle

\medskip 
\medskip

\begin {abstract}

We consider the derivation of  equivalent unconstrained systems for QCD
given in the class of functions of nontrivial topological gauge 
transformations. We show that the unconstrained QCD obtained
by resolving the Gauss law constraint   contains
a monopole, a zero mode of the Gauss law, and a rising potential, 
which can explain the phenomena of confinment and hadronization as well as 
spontaneous chiral symmetry breaking and the $\eta$-$\eta'$-mass 
difference. 
 
\end{abstract} 
 
 
(Key-words: QCD, Gauss law, topology, monopole, zero mode, 
hadronization, confinement, U(1)-problem) 
 
\section{Introduction} 

In this contribution  I discuss the  construction of gauge-invariant physical 
variables in QCD being adequate for the  low energy nonperturbative region.
The validity of the standard Faddeev-Popov (FP) integral~\cite{fp1,fs} for 
arbitrary gauges  in non-Abelian theory has been proved
by Faddeev~\cite{f} only for scattering
amplitudes of degrees of freedom which are not observable in QCD.
This proof is therefore not adequate for description of
 the phenomena of confinement and
hadronization. The FP integral  has been obtained from the Feynman path
integral by construction of
the non-Abelian  equivalent unconstrained system 
in the class of functions of standard
perturbation theory.

In the present paper we discuss the derivation  of equivalent unconstrained
 systems for QCD in the class of functions of topologically nontrivial 
transformations, in order to demonstrate 
that the physical phenomena of hadronization and confinement in QCD 
can be explained by the explicit solutions of the non-Abelian constraints
in this class of functions.

Befor we consider our main idea for QCD, we 
illustrate it on the example of
the equvalent unconstrained system derived
for QED by Heisenberg and Pauli~\cite{hp},
and Dirac~\cite{cj}.

\section{Equivalent Unconstrained Systems in QED} 

Let us consider the action for the Abelian theory
\be \label{qed} 
W=\int d^4x 
\left\{-\frac{1}{4}{G_{\mu \nu}}^2- A^{\mu} J^*_{\mu} \right\}~. 
\ee   
This action contains superfluous (nonphysical)
degrees of freedom and is not compatible
with the simplest variational methods developed
in the framework of the Newtonian mechanics. 

To describe the gauge-invariant dynamics, Heisenberg and Pauli~\cite{hp}, 
and Dirac~\cite{cj} have considered the equivalent system obtained
by resolving the Gauss law constraint. 
Recall that all peculiarities of the application 
of classical variational methods, including the
initial value problem, the spatial boundary conditions,
the time evolution,  the classification 
of constraints, and the equations of motion in the Hamiltonian approach, are
defined only in  a definite frame of reference distinguished by 
the time axis. 

The Gauss law constraint  in the frame of reference with an axis of time 
$l^{(0)}_{\mu}=(1,0,0,0)$ is
the equation for the time component of a gauge field 
\be  \label{glqed} 
\frac{\delta W}{\delta A_0}=0~\Rightarrow~~~\partial_j^2A_0= 
\partial_k \dot A_k +J^*_0~. 
\ee 
As it was shown in Refs.~\cite{hp,cj} (see also~\cite{p2})
the substitution of the solution of this constraint
into the initial action~(\ref{qed}) 
leads to the equvalent unconstrained system 
\be \label{qed*} 
W^*_{l^{(0)}}[A^*,J^*]=\int d^4x 
\left\{-\frac{1}{2}(\partial_{\mu}{A^*_{k}})^2- A^*_i J^*_{i}
-\frac{1}{2}J^*_0(\frac{1}{\partial_j^2}J^*_0)\right\}
~. 
\ee 
By this way, in QED, one could obtain the electrostatic Coulomb interaction
of  sources $J^*_{0}$ in the monopole class of functions 
($f(\vec x)= O(1/r), |\vec x|=r \rightarrow \infty$)
and  two transversal photons 
constructed by Dirac~\cite{cj} as "dressed" variables 
$A^*$ in the explicit form 
\be \label{df} 
ieA^*_k=U(A)(ieA_k+\partial_k )U(A)^{-1}~,~~~~~~~~ 
U(A)=\exp[ie\frac{1}{\partial_j^2 }\partial_kA_k ]~. 
\ee 
 
The equvalent unconstrained system can be quantized by 
the Feynman path integral in the form 
\be \label{qedf} 
Z_F[l^{(0)},J^*] =\int\limits_{ }^{ }d^2A^* 
\exp\bigg\{iW^*_{l^{(0)}} [A^*,J^*]\bigg\} 
\ee  
This path integral depends on the axis of time $l^{(0)}_{\mu}=(1,0,0,0)$.

The relativistic covariance of the equvalent unconstrained system
in QED has been predicted by von Neumann~\cite{hp} and 
proven by Zumino \cite{z} on the level of the algebra of 
generators of the Poincare group. In particular,
a moving relativistic atom in QED is described by the usual boost 
procedure for the wave function, which corresponds to a change of the 
time axis $l^{(0)}\Rightarrow l$, i.e., 
motion of the Coulomb potential~\cite{yaf} itself 
\begin{eqnarray} 
{ W}_{C} 
= \int d^4 x d^4 y \frac{1}{2} 
J^*_{l}(x) V_C(z^{\perp}) 
J^*_{l}(y) \delta(l \cdot z)~~~~(\partial_k^2V_C(\vec x)=-\delta^3(\vec x)), 
\end{eqnarray} 
where 
$ 
J^*_{l} = l^{\mu} J^*_{\mu} \,\, , \,\, 
z_{\mu}^{\perp} = 
z_{\mu} - l_{\mu}(z \cdot l) \,\, , \,\, 
z_\mu = (x - y)_\mu  \,\, . \,\, 
$ 
A time axis is chosen 
to be parallel to the total momentum of a considered state.
In particular, for bound states this choice means that the 
coordinate of the potential coincides with the space of the relative 
coordinates of the bound state wave function in the accordance
with  the Yukawa-Markov principle~\cite{ym} and the Eddington concept of 
simultaneity ("yesterday's electron and today's proton do not make 
an atom")~\cite{ed}. In this case,
we get the relativistic covariant dispersion law and invariant mass spectrum.  
The  relativistic generalization of the Coulomb potential  is not only 
the change of the form of the potential, but also the change of a 
direction of its motion in four-dimensional space to lie along the total 
momentum of the bound state. 
The relativistic covariant unitary perturbation theory 
in terms of such the relativistic 
instantaneous bound states has been constructed in~\cite{yaf}. 
In this pertusbation theory, each instantaneous bound state in QED has a 
proper equvalent unconstrained system. The manifold of frames corresponds to 
the manifold of "equivalent unconstrained systems". 
The relativistic invariance means that a complete set of physical 
states for any equivalent system coincides with the one for another 
equivalent system (this treatment belongs to Schwinger, see~\cite{bww}).   

This treatment of the relativistic invariance and covariance
is confused with 
the naive  understanding of the relativistic invariance as 
independence on the time-axis of the functional integral. 
  
One supposes that the dependence on the frame ($l^{(0)}$) can be removed 
by the transition from the Feynman integral~(\ref{qedf}) 
to perturbation theory in any relativistic-invariant 
gauge $f(A)=0$ with the FP determinant 
\be \label{qedr} 
Z_{FP}[J] =\int\limits_{ }^{ }d^4A \delta[f(A)]\Delta_{FP} 
\exp\bigg\{iW[A,J]\bigg\}~. 
\ee 
This transition is well-known as a "change of gauge", and it 
is fulfilled in two steps \\ 
I) the change of the variables $A^*$~(\ref{df}), and \\ 
II) the change of the physical sources $J^*$ of the type of 
\be \label{ps} 
A^*_k(A)J_k^*=U(A)\left(A_k-\frac{i}{e}\partial_k \right)U^{-1}(A) J^*_k 
~\Rightarrow~A_{k}J^{k}~. 
\ee 
At the first step, all electrostatic monopole physical phenomena 
are concentrated in the Dirac gauge factor $U(A)$~(\ref{df}) 
that accompanies the physical sources $J^*$. 
 
At the second step, changing the sources~(\ref{ps}) we lose the Dirac factor 
together with the whole class 
of electrostatic phenomena including the Coulomb-like instantaneous bound 
state formed by the electrostatic interaction. 
 
Really, the FP perturbation theory in the relativistic gauge contains only 
photon propagators with the light-cone singularities forming 
the Wick-Cutkosky bound states with the spectrum  differing 
\footnote{The author thanks W. Kummer who pointed out that in 
Ref. \cite{kum} the difference between the Coulomb atom and 
the Wick-Cutkosky bound states in QED has been demonstrated.} 
from the observed one which corresponds to the instantaneous Coulomb 
interaction. 
Thus, the restoration of the ``explicit'' relativistic form of
the equvalent unconstrained system ($l^{(0)}$) 
by the transition to a relativistic gauge 
loses all electrostatic "monopole physics" with the Coulomb bound states.

\section{Unconstrained QCD} 
 
\subsection{Topological degeneration and class of functions} 
 
We consider the non-Abelian $SU_c(3)$ theory with the action functional 
\be \label{u} 
W=\int d^4x 
\left\{\frac{1}{2}({G^a_{0i}}^2- {B_i^a}^2) 
+ \bar\psi[i\gamma^\mu(\partial _\mu+{\hat 
A_\mu}) 
-m]\psi\right\}~, 
\ee 
where $\psi$ and $\bar \psi$ are the fermionic quark fields. 
We use the conventional notation for the non-Abelian electric 
and magnetic fields 
\be \label{v} 
G_{0i}^a = \partial_0 A^a_i - D_i^{ab}(A)A_0^b~,~~~~~~ 
B_i^a=\epsilon_{ijk}\left(\partial_jA_k^a+ 
\frac g 2f^{abc}A^b_jA_k^c\right)~, 
\ee 
as well as the covariant derivative 
$D^{ab}_i(A):=\delta^{ab}\partial_i + gf^{acb} A_i^c$. 
 
The action (\ref{u}) is invariant with respect to gauge transformations 
$u(t,\vec x)$ 
\be \label{gauge1} 
{\hat A}_{i}^u := u(t,\vec x)\left({\hat A}_{i} + \partial_i 
\right)u^{-1}(t,\vec x),~~~~~~ 
\psi^u := u(t,\vec x)\psi~, 
\ee 
where ${\hat A_\mu}=g\frac{\lambda^a }{2i} A_\mu^a~$. 
 
It is well-known~\cite{fs} that the initial data of all fields are 
degenerated with respect to the stationary gauge transformations 
$u(t,\vec{x})=v(\vec{x})$. 
The group of these transformations represents 
the group of three-dimensional paths lying  on the three-dimensional 
space of the $SU_c(3)$-manifold with the homotopy group 
$\pi_{(3)}(SU_c(3))=Z$. 
The whole group of stationary gauge transformations is split into 
topological classes marked by the integer number $n$ (the degree of the map) 
which counts how many times a three-dimensional path turns around the 
$SU(3)$-manifold when the coordinate $x_i$ runs over the space where it is 
defined. 
The stationary transformations $v^n(\vec{x})$ with $n=0$ are called the small 
ones; and those with $n \neq 0$ 
\be \label{gnl} 
{\hat A}_i^{(n)}:=v^{(n)}(\vec{x}){\hat A}_i(\vec{x}) 
{v^{(n)}(\vec{x})}^{-1} 
+L^n_i~,~~~~L^n_i=v^{(n)}(\vec{x})\partial_i{v^{(n)}(\vec{x})}^{-1}~, 
\ee 
the large ones. 
 
The degree of a map 
\be \label{gn2} 
{\cal N}[n] 
=-\frac {1}{24\pi^2}\int d^3x ~\epsilon^{ijk}~ Tr[L^n_iL^n_jL^n_k]=n~. 
\ee 
as the condition for 
normalization  means that the large transformations 
are given in the  class of functions with the spatial asymptotics 
${\cal O} (1/r)$. 
Such a function $L^n_i$~(\ref{gnl}) is given by 
\be \label{class0} 
v^{(n)}(\vec{x})=\exp(n \hat \Phi_0(\vec{x})),~~~~~ 
\hat \Phi_0=- i \pi\frac{\lambda_A^a x^a}{r} f_0(r)~, 
\ee 
where the antisymmetric SU(3) matrices are denoted by 
$$\lambda_A^1:=\lambda^2,~\lambda_A^2:=\lambda^5,~\lambda_A^3:=\lambda^7~,$$ 
and $r=|\vec x|$. 
The function $f_0(r)$ satisfies the boundary conditions 
\be \label{bcf0} 
f_0(0)=0,~~~~~~~~~~~~~~ 
f_0(\infty)=1~, 
\ee 
so that the functions $L_i^n$ disappear at spatial infinity 
$\sim {\cal O} (1/r)$. 
The functions $L_i^n$ belong to monopole-type class of 
functions. It is evident that the transformed physical fields $A_i$ 
in~(\ref{gnl}) should be given in the same class of functions
$\hat {\Phi}_i(\vec{x})= O(\frac{1}{r})$ 
\be \label{bar} 
A^c_i(t,\vec{x}) = {\Phi}^c_i(\vec{x}) + \bar A^c_i(t,\vec{x})~, 
\ee 
where $\bar A_i$ is a weak perturbative part 
with the asymptotics at the spatial infinity 
\be \label{ass1} 
\bar A_i(t,\vec{x})|_{\rm asymptotics} = O(\frac{1}{r^{1+l}})~~~~(l > 1)~. 
\ee 
We restrict ouselves to 
ordinary perturbation theory around a static monopole $\Phi_i(\vec x)$, and 
use, as an example, the Wu-Yang monopole~\cite{wy,fn} 
\be \label{wy} 
\Phi_i^{WY}= 
- i \frac{\lambda_A^a}{2}\epsilon_{iak}\frac{x^k}{r^2} f_1^{WY},~~~~~~~ 
f^{WY}_{1}=1 
\ee 
which is a solution of classical equations everywhere besides 
the origin of coordinates. 
To remove a sigularity at the origin of coordinates and regularize 
its energy, the Wu-Yang monopole is changed by the 
Bogomol'nyi-Prasad-Sommerfield (BPS) monopole~\cite{bps} 
\be \label{bps}
f^{WY}_{1}~\Rightarrow~ 
f^{BPS}_{1}= 
\left[1 - \frac{r}{\epsilon \sinh(r/\epsilon)}\right]~,~~~~~ 
\int\limits_{ }^{ }d^3x [B^a_i(\Phi_k)]^2 =\frac{4\pi}{g^2 \epsilon}~, 
\ee 
to take the limit of zero size $\epsilon~\rightarrow~ 0$ at the 
end of the calculation of spectra and matrix elements.

The statement of the problem is to construct an 
equivalent unconstrained system  for the non-Abelian fields 
in this monopole-type class of functions. 
 
\subsection{The Gauss Law Constraint} 
 
An equivalent unconstrained system is obtained by resolving
the non-Abelian Gauss law constraint  
\be \label{gaussd} 
\frac{\delta W}{\delta A_0}=0~~~~~ \Rightarrow 
(D^2(A))^{ac} { A_0}^c = D_i^{ac}(A)\partial_0 A_i^c+ j_0^a~, 
\ee 
where $j_\mu^a=g\bar \psi \frac{\lambda^a}{2} \gamma_\mu\psi$ 
is the quark current. 

In lowest order of the perturbation theory this constraint takes the form
\be \label{gausspt} 
(D_j^2(\Phi))^{ac} { A_0}^c = D_i^{ac}(\Phi)\partial_0 A_i^c~, 
\ee  
In the considered case of BPS-monopole (\ref{bps}), there is
the zero mode of the covariant Laplace operator in the monopole field 
\be \label{lap} 
(D^2)^{ab}({\Phi_k^{BPS}})({\Phi}_0^{BPS})^b(\vec{x})=0~. 
\ee 
The nontrivial solution of this equation is well-known~\cite{bps}; 
it is given by equation~(\ref{class0}) where 
\be \label{lapb} 
f_0^{BPS}=\left[ \frac{1}{\tanh(r/\epsilon)}-\frac{\epsilon}{r}\right] 
\ee 
has the boundary conditions~(\ref{bcf0}) of a phase of the 
topological transformations~(\ref{class0}). 

This zero mode is a perturbative form of a zero mode of the Gauss law 
constraint~(\ref{gaussd})~\cite{p2,vp1,n} as the solution 
${\cal Z}^a$  of the homogeneous equation~(\ref{gaussd}) 
\be\label{zm} 
(D^2(A))^{ab}{\cal Z}^b=0~, 
\ee 
with the asymptotics at the space infinity 
\be \label{ass} 
\hat {\cal Z}(t,\vec{x})|_{\rm asymptotics}=\dot N(t)\hat \Phi_0(\vec{x})~, 
\ee 
where $\dot N(t)$ is the global variable of an excitation 
of the gluon system as a whole.
From the mathematical point of view, the zero mode means that 
the general solution of the inhomogeneous equation~(\ref{gaussd}) 
for the time-like component $A_0$ 
is a sum of the homogeneous equation~(\ref{zm}) 
and a particular solution 
${\tilde A}_0^a$ of the inhomogeneous one~(\ref{gaussd}): 
$A_0^a = {\cal Z}^a + {\tilde A}^a_0$~. 
 
The zero mode of the Gauss constraint and the 
topological variable $N(t)$ allow us to remove the topological 
degeneration of all fields by the non-Abelian generalization of 
the Dirac dressed variables~(\ref{df}) 
\be \label{gt1} 
0=U_{\cal Z}(\hat {\cal Z}+\partial_0)U_{\cal Z}^{-1}~,~~~~ 
{\hat A}^*_i=U_{\cal Z}({\hat A}^I+\partial_i)U_{\cal Z}^{-1},~~~ 
\psi^*=U_{\cal Z}\psi^I~, 
\ee 
where the spatial asymptotics of $U_{\cal Z}$ is 
\be \label{UZ} 
U_{\cal Z}=T\exp[\int\limits^{t} dt' 
\hat {\cal Z}(t',\vec{x})]|_{\rm asymptotics} 
=\exp[N(t)\hat \Phi_0(\vec{x})]=U_{as}^{(N)}~, 
\ee 
and $A^I=\Phi+\bar A,\psi^I$ are the degeneration free variables.
In this case, the topological degeneration of all color fields 
converts into the degeneration of only one global topological 
variable $N(t)$ with respect to a shift of this variable on integers: 
$(N~\Rightarrow~ N+n,~ n=\pm 1,\pm 2,...)$. 
One can check~\cite{bpr} that the Pontryagin index for 
the Dirac variables~(\ref{gt1}) with the 
assymptotics~(\ref{ass1}),~(\ref{ass}),~(\ref{UZ}) is determined 
by only the diference of the final and initial values of 
the topological variable 
\be \label{pont} 
\nu[A^*]=\frac{g^2}{16\pi^2}\int\limits_{t_{in} }^{t_{out} }dt 
\int\limits_{ }^{ }d^3x G^a_{\mu\nu} {}^*G^{a\mu\nu}=N(t_{out}) -N(t_{in})~. 
\ee 
Thus, we can identify the global variable $N(t)$ with
the winding number degree of freedom in the Minkowski space. This degree
of freedom plays the role of the Goldstone mode that removes the topological
degeneration of initial data. 

As we shall see below, the vacuum wave function of the topological free
 motion 
in terms of the Ponryagin index~(\ref{pont}) takes 
the form of a plane wave $\exp(i P_N \nu[A^*])$. 
The well-known instanton wave 
function~\cite{hooft} appears for nonphysical values of the topological 
momentum $P_N=\pm i 8 \pi^2/g^2$.
This fact points out that instantons are permanently tunelling
and correspond to quantun
 nonphysical solutions with the zero energy 
in Euclidean space-time of the type of nonphysical ones for
an oscillator $(\hat p^2+q^2)\psi_0=0$
\footnote{ The author is grateful to V.N. Gribov for the discussion of 
the problem of instantons during a visit in  Budapest,
May 1996.}.

\subsection{Physical Consequences} 
 
The dynamics of physical variables including the topological one 
is determined by the constraint-shell action of an equivalent unconstrained 
system (EUS) as a sum 
of the zero mode part, and the monopole and perturbative ones 
\be \label{csa} 
W^*_{l^{(0)}}=W_{Gauss-shell}=W_{\cal Z}[N]+W_{mon}[\Phi_i]+W_{loc}[\bar A]~. 
\ee 
The action for an equivalent unconstrained system~(\ref{csa}) in the 
gauge~(\ref{qcdg}) with a monopole and a zero mode has been 
obtained  in the paper~\cite{bpr} following the paper~\cite{f}. 
This action contains the dynamics of the topological variable in the 
form of a free rotator 
\be \label{ktg} 
W_{\cal Z}=\int\limits_{ }^{ }dt\frac{{\dot N}^2 I}{2};~~~ 
I=\int\limits_{V}d^3x(D^{ac}_i(\Phi_k)\Phi^c_0)^2= 
\frac{4\pi}{g^2}(2\pi)^2\epsilon~, 
\ee 
where $\epsilon$ is a size of the BPS monopole considered 
as a parameter of the infrared regularization which disappears 
in the infinite volume limit. The dependence of $\epsilon$ on 
volume can be chosen as $\epsilon \sim V^{-1}$, so that the density of energy
 was finite.

The perturbation theory in the sector of local excitations $W_{loc}[\bar A]$
is based  on the Green function as the inverse 
differential operator of the Gauss law 
\be \label{V}
[D^2(\Phi)]^{ac}V^{cb}(x,y)=-\delta^3(x-y)\delta^{ab}
\ee 
which is the non-Abelian generalization of the Coulomb potential. 
As it has been shown in~\cite{bpr}, the non-Abelian Green function (\ref{V})
in the field of the Wu-Yang monopole 
is the sum of a Coulomb-type potential and a rising one. 
This means that the instantaneous quark-quark interaction 
leads to spontaneous chiral symmetry breaking~\cite{yaf,fb}, 
goldstone mesonic bound states~\cite{yaf}, glueballs~\cite{fb,ac}, and 
the Gribov modification of the asymptotic freedom formula~\cite{ac}. 
If we choose a time-axis $l^{(0)}$ 
along the total momentum of bound states~\cite{yaf} 
(this choice is compatible with the experience of QED in the description 
of instantaneous bound states), we get the bilocal generalization of 
the chiral Lagrangian-type mesonic interactions~\cite{yaf}. 
In this case, the U(1) anomalous interaction of $\eta_0$-meson with
the topological variable~\cite{bpr} lead to additional mass of this 
isoscalar meson. 

All these results can be described by  
the Feynman path integral for the obtained unconstrained system in 
the class of functions of the topological transformations (see~\cite{bpr}) 
\bea \label{qcdf} 
Z_F[l^{(0)},J^{a*}]&=&\int\limits_{ }^{ } DN(t) 
\int\limits_{ }^{ }\prod\limits_{c=1 }^{c=8 } 
[d^2A^{c*} d^2 E^{c*}]\nonumber\\ 
&&\times\exp\left\{iW^*_{l^{(0)}} [A^*,E^*]+i\int\limits_{ }^{ }d^4x 
[J^{c*}_{\mu} \cdot A^{c*}_{\mu}]\right\}~, 
\eea 
where $J^{c*}$ are physical sources. 
 
The considered case of the Wu-Yang monopole corresponds 
to the factorization of the phase factors of the topological 
degeneration, so that 
the physical consequences of the degeneration with respect to the 
topological nontrivial initial data are determined by the gauge of 
the sources of the Dirac dressed fields $A^*,\psi^*$ 
\be \label{tcsa} 
W^*_{l^{(0)}}(A^*) + \int\limits_{ }^{ }d^4x J^{c*}A^{c*}= 
W^*_{l^{(0)}}(A^I) +\int\limits_{ }^{ } d^4x J^{c*}A^{c*}(A^I)~. 
\ee 
The nonperturbative 
phase factors of the topological degeneration can lead to 
a complete destructive interference of color amplitudes~\cite{p2,n,pn} 
due to averaging over all parameters of the degenerations, in particular 
\be \label{conf} 
<1|\psi^*|0>=<1|\psi^I|0> \lim\limits_{L \to \infty} 
\frac{1}{2L} \sum\limits_{n=-L }^{n=+L } U_{as}^{(n)}(x)=0~. 
\ee 
This mechanism of confinement due to the interference 
of phase factors (revealed by the explicit 
resolving the Gauss law constraint~\cite{n}) disappears 
after the change of "physical" sources $A^*J^*~\Rightarrow~A J$ that 
is called the transition to another gauge. In the lowest order of
perturbation theory the Gauss law constraint for the degeneration free 
variables is compatible with only one gauge.
 It is the Coulomb-type gauge in the monopole field 
\be \label{qcdg} 
D_k^{ac}(\Phi)\bar A^c_k=0~. 
\ee 
 
The change of variables $A^*$ of the type of~(\ref{df}) 
with the non-Abelian Dirac factor 
\be \label{dirqcd} 
U(A)=U_{\cal Z}\exp\left\{\frac{1}{D^2(\Phi)} D_j(\Phi)\hat A_j\right\} 
\ee 
and the change of the Dirac sources $J^*$ can remove all 
monopole physics, including confinement and hadronization, 
like similar changes~(\ref{df}),~(\ref{ps}) in QED (to get a 
relativistic form of the Feynman path integral) 
remove all electrostatic phenomena in the relativistic gauges. 
 
The transition to another gauge faces the problem of zero 
of the FP determinant $det D^2(\Phi)$ (i.e. the Gribov ambiguity~\cite{g} of 
the gauge~(\ref{qcdg})). It is the zero mode of the second class 
constraint. The considered example~(\ref{qcdf}) shows that 
the Gribov ambiguity (being simultaneously the zero mode of the first 
class constraint) cannot be removed by the change of gauge 
as the zero mode is the inexorable consequence of internal dynamics, like the 
Coulomb field in QED. Both the zero mode, in QCD, and the Coulomb field, 
in QED, have nontrivial physical consequences discussed above, 
which can be lost by the standard gauge-fixing scheme.

\section{Conclusion} 
 
There are
``admissible'' gauges for which the operations of varying and constraining
commute~\cite{f,cj,rev}.
 This commutativity allows us to construct an equivalent
unconstrained system (compatible with the Feynman path integral)
directly in terms of gauge-invariant variables
by using the substitution of a solution of the Gauss law constraint into
the initial singular action. 

In this paper we reproduced the Faddeev derivation of the non-Abelian
unconstrained system in another class of functions.
This class of functions is unambigiously
defined by the normalization of nontrivial topological gauge
transformations and contains a monopole and a zero mode of
the Gauss law constraint. 

We have shown that "equivalent unconstrained systems" in QCD
in the class of functions of nontrivial topological transformations
contains a monopole and a zero mode of the Gauss law constraint.
The monopole forms the rising potential of hadronization of
color quarks and gluons. And the zero mode forms the phase factors
of the topological degeneration and additional mass of the $\eta_0$ -
meson.

There is only one gauge
for which the operations of varying and constraining
commute. It is not Coulomb one, but its
the covariant generalization in the presence of a monopole. 

If we pass to another gauges on the level of the FP integral
in relativistic gauges, all these monopole phenomena  are lost.

Recall that Faddeev proved the equivalence of the Feynman integral 
to the Faddeev-Popov integral in an arbitrary gauge
 for the scattering amplitudes~\cite{f} only, i.e. when  
all particle-like excitations of the fields are on their mass-shell. 
However, for the cases of bound states in QED and QCD
 and  other collective phenomena where these fields are off their mass-shell 
this equivalence has not been proved and might not exist.

\section*{Acknowledgments} 
 
\medskip 
 
I thank  Profs. D. Blaschke, J. Polonyi, and G. R\"opke 
for critical discussions. This research and the participation at the 
workshop ``Quark matter in Astro- and Particle Physics'' was supported
in part by funds from the Deutsche Forschungsgemeinschaft and the Ministery 
for Education, Science and Culture in Mecklenburg- Western Pommerania.

\end{document}